# BELL INEQUALITIES VIOLATION IN RELATIVITY THEORY

*A.V.Belinsky[1], I.I.Dzhadan[2].*

*Faculty of Physics, Lomonosov Moscow State University*

**Annotation**. A violation of Bell local realism inequalities in Clauser-Horn-Shimony-Holt (CHSH) form has been discovered in a relativistic GedanknExperiment. This means that there are no definite joint probabilities and this finds a classical explanation in the structure of Special relativity (SRT). The discovered nonlocality is weaker than Albert Einstein's quantum 'spooky action at a distance', and its presence suggests certain parallels between special relativity and quantum mechanics.

**Keywords**: Bell inequalities, quantum nonlocality, quantum superposition, statistical independence, SRT, 'spooky action at a distance'.

## Introduction

Since the realization of the problem of objective reality in quantum mechanics, associated with the publication of the Einstein-Podolsky-Rosen article [1,2], attempts to explain this phenomenon have continued. A similar problem has not been found yet in non-quantum physics. However, a simple mathematical consideration suggests that if the number of random variables is greater than three, then under certain conditions there may be a case that the classical probability distribution for these variables cannot be constructed [3,4], in Special relativity theory (SRT) for example.

## Bell-Clauser-Horn-Shimony-Holt inequalities (CHSH)

Before examining the CHSH inequalities in SRT, let's consider which physical assumptions are refuted if CHSH are violated in statistical experiments. Let us briefly recall the derivation of the Bell inequality [5] in the form of CHSH [6] with the emphasis relevant to our situation. In the simplest case [7,8] the arithmetic formula $s_i = a_i(b_i + b_i') + a_i'(b_i - b_i')$ is taken where $a_i^{(\prime)}, b_i^{(\prime)} = \pm 1$ are the values of dichotomous random variables obtained from 4 single-bit measurements, and combined into one '4-dimensional' variable with index $i$. $s_i$ can only take two values: $+2$ and $-2$ as one term in parentheses is equal to 0, hence $|s_i| = 2$, and given multiplication by probabilities to obtain averages, leads to the inequality $|s_i| \leq 2$.

It is implying that $a_i^{(\prime)}, b_i^{(\prime)}$ are mutually independent. Otherwise if there were $s_i = a_i\big(b_{i(a_i)} + b_{i(a_i)}'\big) + a_i'(b_{i(a_i')} - b_{i(a_i')}')$ with $b_{i(a_i)}^{(\prime)} \neq b_{i(a_i')}^{(\prime)}$ the inequality $|s_i| \leq 2$ could be violated. Thus, the inequality $|s_i| \leq 2$ is universally true if the following conditions hold: the coexistence of values in

---
[1] E-mail: belinsky@physics.msu.ru
[2] E-mail: idzhadan@yandex.ru

each 4-dimensional realization, and the number of values must be equal to 4, since in the case of $b_{i(a_i)}^{(\prime)} \neq b_{i(a_i')}^{(\prime)}$ this number becomes greater and the inequality could be violated on this ground.

The coexistence of values is generally treated as 'realism', and the requirement of $b_{i(a_i)}^{(\prime)} = b_{i(a_i')}^{(\prime)}$ is interpreted, as 'locality'. Therefore, the joint grounds for the inequality derivation can be the physical necessities of locality and realism. According to de Morgan's logical rule, the negation of inequality implies the denial of locality or realism, or both. On this basis, it is usual in physical language to speak about the local realism violation as a "physical cause" of the violation of the corresponding inequalities, although the violation of local realism does not necessarily imply the violation of inequalities. On the contrary: the violation of inequalities leads to the violation of locality, or realism, or both.

If the described four-dimensional measurement is repeated a sufficiently large number of times N≫1, the following inequality will be true:

$$S_i = \left|\sum_{i=1}^{N} s_i\right| = |a_i(b_i + b_i') + a_i'(b_i - b_i')| = |a_i b_i + a_i b_i' + a_i' b_i - a_i' b_i'| \leq 2N \quad (1)$$

Let's denote the variants of realisations of $a_i$ and $b_i$ in (1) as (++),(— —),(+—), (—+) and the number of occurrences of each variant in a series of N 4-dimensional measurements for the values of $a_i^{(\prime)}$ and $b_i^{(\prime)}$ for random variables $A^{(\prime)}$ and $B^{(\prime)}$: $n_{a^{(\prime)}b^{(\prime)}}^{++}, n_{a^{(\prime)}b^{(\prime)}}^{--}, n_{a^{(\prime)}b^{(\prime)}}^{+-}, n_{a^{(\prime)}b^{(\prime)}}^{-+}$. Then:

$$n_{a^{(\prime)}b^{(\prime)}}^{++} + n_{a^{(\prime)}b^{(\prime)}}^{--} + n_{a^{(\prime)}b^{(\prime)}}^{+-} + n_{a^{(\prime)}b^{(\prime)}}^{-+} = N.$$

Let's denote the number of correlations and the number of anticorrelations for the values of a pair of random variables obtained when measuring $A^{(\prime)}$ and $B^{(\prime)}$:

$$N_{a^{(\prime)}b^{(\prime)}}^{cor} \stackrel{\text{def}}{=} n_{a^{(\prime)}b^{(\prime)}}^{++} + n_{a^{(\prime)}b^{(\prime)}}^{--}, N_{a^{(\prime)}b^{(\prime)}}^{ant} \stackrel{\text{def}}{=} n_{a^{(\prime)}b^{(\prime)}}^{+-} + n_{a^{(\prime)}b^{(\prime)}}^{-+}.$$

Let us define the correlation function for a pair of random variables $A^{(\prime)}$ and $B^{(\prime)}$ for a sufficiently large N as:

$$\langle A^{(\prime)} B^{(\prime)} \rangle \stackrel{\text{def}}{=} \frac{\left(N_{a^{(\prime)}b^{(\prime)}}^{cor} - N_{a^{(\prime)}b^{(\prime)}}^{ant}\right)}{N} = P_{A^{(\prime)}B^{(\prime)}}^{cor} - P_{A^{(\prime)}B^{(\prime)}}^{ant}$$

Since the product $a_i^{(\prime)} b_i^{(\prime)}$ is always equal to +1 for correlated pairs and is always equal to –1 for anticorrelated pairs, then:

$$\langle A^{(\prime)} B^{(\prime)} \rangle = \frac{\left(\sum_{i=1}^{N} a_i^{(\prime)} b_i^{(\prime)}\right)}{N} = P_{A^{(\prime)}B^{(\prime)}}^{cor} - P_{A^{(\prime)}B^{(\prime)}}^{ant}$$

and inequality (1) turns into CHSH

$$|\langle AB \rangle + \langle AB' \rangle + \langle A'B \rangle - \langle A'B' \rangle| \leq 2. \tag{2}$$

It can also be derived from the existence of four-dimensional classic probability distributions [9]. In some cases, it may be more convenient to use an equivalent inequality for probabilities. Since for a sufficiently large $N$ $P^{cor}_{A^{(\prime)}B^{(\prime)}} = \frac{N^{cor}_{a^{(\prime)}b^{(\prime)}}}{N}$ and $P^{ant}_{A^{(\prime)}B^{(\prime)}} = \frac{N^{ant}_{a^{(\prime)}b^{(\prime)}}}{N}$, then:

$$\langle A^{(\prime)} B^{(\prime)} \rangle = P^{cor}_{A^{(\prime)}B^{(\prime)}} - P^{ant}_{A^{(\prime)}B^{(\prime)}} = 2P^{cor}_{A^{(\prime)}B^{(\prime)}} - 1,$$

and (2) can be rewritten as:

$$0 \leq P^{cor}_{AB} + P^{cor}_{AB'} + P^{cor}_{A'B} - P^{cor}_{A'B'} \leq 2. \tag{3}$$

### GedankenExperiment with four siblings

A quadruplet embarked on a space journey in 4 spaceships travelling at different speeds. Let's designate these siblings $R_1$, $R_2$, $R_3$, $R_4$. After a while, one of them ($R_1$) decided to send messages to other siblings, however, he did not know if his brothers were still alive. All he knew was that the probability $P(S_j)$ of the twin $R_j$ death in the time span $\Delta t_j$ measured in his own inertial frame $S_j$ ($j \in \{1,2,3,4\}$) was $P(S_j)=k\Delta t_j$, where $k$ is the known family probability coefficient of a twin's death per unit of time. Before starting the correspondence, the twin decided to calculate the probability that his brother was alive for each of him. The twin knew physics, and understood that if the proper frame $S_i$ of the twin $R_i$ was defined as 'moving' and the $S_j$ of the twin $R_j$ as 'rest', then the clock in $S_i$ would lag behind the clock in $S_j$ as a result of relativistic 'time dilation'

$$\Delta t_i = \Delta t_j / \gamma = \Delta t_j \sqrt{1 - \frac{v_{ij}^2}{c^2}}, \ v_{ij} - \text{the velocity of } R_i \text{ relative to } R_j.$$

Thus, from the $R_j$ twin point of view the probability of a random event occurring per unit of the $R_i$ twin's own time $\Delta t_i$ with the body that is moving with $S_i$ will be $\gamma$ times less:

$$P(A_i) = P(A_j)/\gamma = P(A_j)\sqrt{1 - \frac{v_{ij}^2}{c^2}} \stackrel{\text{def}}{=} r_{ij} P(A_j), 0 < r_{ij} \leq 1. \tag{4}$$

Using equation (4) a twin can calculate the probabilities that his brother is alive at the moment, and knowing their mutual distances, he also can calculate the probability that his brother would still be alive at the time of receiving his message. But the twin decided not to limit himself to this.

On the basis of (4), the twin calculated the conditional probability $P^{+|-}_{ij}$ that the twin $R_i$ had already died (+), assuming that the twin $R_j$ is still alive (–). Similarly, he also calculated the conditional probability $P_{ij}^{-|-}$ that the twin $R_i$ was alive (–) assuming that the $R_j$ twin was alive (–) too for two arbitrary twins that were moving relative to each other at a speed of $\pm v_{ij}$

$$P_{ij}^{+|-} = r_{ij} k \Delta t,$$

$$P_{ij}^{-|-} = r_{ij}(1 - k\Delta t), \text{ were } r_{ij} = \sqrt{1 - \frac{v_{ij}^2}{c^2}} \qquad (5)$$

To enable any of the four twins to assess the probability of recipient being alive at the time of arrival of the message and the observer being alive at the time of arrival of the response, the first twin calculated pairwise correlations for the life and death of the twins. Since the conditional probabilities $P_{ij}^{-|+}$ and $P_{ij}^{+|+}$ are taken under the condition of the death of the observer twin and are therefore meaningless, the actual probability distribution contains only $P_{ij}^{+|-}$ and $P_{ij}^{-|-}$. Then

$$P_{ij}^{cor} = P_{ij}^{-|-} = r_{ij}(1 - k\Delta t) \qquad (5a)$$

The coordinate velocities of the four twins which in $S_2$ amounted to:

$$v_1 = -0.4c, v_2 = 0, v_3 = 0, v_4 = 0.4c.$$

Using the formula of relativistic addition of velocities [10,11]

$$v_{ij} = \frac{v_i + V_{ji}}{1 + \frac{v_i V_{ji}}{c^2}}, V_{ji} = -v_j,$$

a twin was able to find the relative velocities

$$v_{12} = -0.4c, v_{23} = 0, v_{34} = -0.4c, v_{14} \approx -0.6897$$

Next, he calculated the necessary coefficients using the formula $r_{ij} = \sqrt{\left(1 - \left(\frac{v_{ij}}{c}\right)^2\right)}$, obtaining: $r_{12} = r_{34} \approx 0.9165, r_{23} = 1, r_{14} \approx 0.7240$. Since for a pair of $A_i$ and $A_j$ the random variables are $\langle A_i A_j \rangle \stackrel{\text{def}}{=} P_{A_i A_j}^{cor} - P_{A_i A_j}^{ant} = 2P_{A_i A_j}^{cor} - 1$, the correlations were calculated as follows:

$$\langle A_i A_j \rangle = 2P_{ij}^{cor} - 1 = 2P_{ij}^{-|-} - 1 = 2r_{ij}(1 - k\Delta t) - 1 \qquad (6)$$

In addition to relativistic formulas, the first twin knew the CHSH inequality, which, in accordance with the order of designation of twins, looks like this

$$|\langle A_1 A_2 \rangle + \langle A_2 A_3 \rangle + \langle A_3 A_4 \rangle - \langle A_1 A_4 \rangle| \leq 2 \text{ — for the correlation function and}$$

$$0 \leq P_{12}^{cor} + P_{23}^{cor} + P_{34}^{cor} - P_{14}^{cor} \leq 2 \text{ — for probabilities} \qquad (7)$$

Given (5a) and (6)

$$P_{12}^{cor} + P_{23}^{cor} + P_{34}^{cor} - P_{14}^{cor} = (r_{12} + r_{23} + r_{34} - r_{14})(1 - k\Delta t).$$

Substituting the necessary relativistic coefficients $r_{12} = r_{34} \approx 0.9165$, $r_{23} = 1, r_{14} \approx 0.7240$, the twin obtained the following inequality for the probabilities $S \approx 2.1090(1 - k\Delta t) \leq 2$, which is violated at $1 - k\Delta t > \frac{2}{2.1090} \approx 0.9483$, that is at the probability of the twin's death $k\Delta t < 0.0517$. Under the same conditions, the inequality for correlations is also violated.

The twin had to conclude that it was impossible to establish common conditions for all brothers to start correspondence, since there was no classical joint probability distribution. The violation of CHSH also implies the negation of life and death realism, thus the siblings are not even able to come to a consensus on whether their brother is alive or not.

## Conclusion

The result seems surprising, since the siblings' death events were defined as an independent and random. In spite of this, CHSH is violated with a certain choice of parameters. The CHSH inequality is never violated with 4-dimensional probability distributions, since it is inferred from the existence of such distributions. According to de Morgan's logical rule, the violation of CHSH derived from the joint premises leads to the violation of at least one of the premises. Thus, a violation of the CHSH inequality means that either the joint probability distribution does not exist in principle, or it is more than 4-dimensional.

In quantum mechanics, the violation of the CHSH clearly follows from the formalism of the theory. However, in case of SRT, quantum entanglement and superposition are absent. Nevertheless, it turns out that twins cannot establish the probability of the brothers' death unambiguously. They are not even able to come to a consensus on whether their brother is alive or not. What is the mechanism of the local realism violation in relativistic theory?

The physical reason for the ambiguity of joint properties in special relativity is apparently associated with the absence of common present, which would make it possible to unambiguously determine the 'jointness' of the events. Suppose there are two distant worldlines of twins, each with a point corresponding to the death event. Let these world lines be marked with – before death, and + after death. A 'present' is a 3-dimensional hyperplane in 4-dimensional space-time that intersects the world lines. Depending on the relative velocity, we must consider a different hyperplane of 'present' for specific sibling. Consequently, the 'present' can intersect the two world lines in different ways on the sections of the world lines marked as (– +), (– –), (+ –), (+ +). The choice of relative velocity will determine how long during the experiment the 'present' of specific sibling will belong to one or another of these four options: if longer, the probability of this occurrence, calculated for many homogeneous experiments with twins, will be greater; if shorter, then less. Accordingly, the correlations will be drawn from different distributions belonging to different 'presents' which leads to the violation of the local realism inequalities. The properties now depend on relative velocities.

Apparently, such a relationship between distant events in special relativity is due to a nonlocal definition of the 'relative velocity' concept. It is given as a binary relationship between distant bodies, although such 'nonlocality' does not indicate the presence of any physical interaction between the bodies. All physical bodies, so to speak, are 'classically entangled' with each other in terms of relative velocities, in the sense that if the relative velocity of body $A$ with respect to body $B$ is equal to $v$, then the relative velocity of body $B$ with respect to body $A$ is strictly equal to $-v$.

As we can see, this 'entanglement' is rather absolute than statistical; if we "can predict" one speed, we know the other with absolute accuracy and reliability. From this, it is clear that when relative velocity indirectly influences joint probabilities by determining the 'present', the observed effect may appear to be 'instantaneous action at a distance', although it is not. This kind of 'nonlocality' should not be confused with Einstein's 'spooky action at a distance'. This 'nonlocality' stems from the nonlocal rules for the distribution of vector quantities (relative velocities) in spacetime and is obviously a weaker physical assumption than 'instantaneous action at a distance'. Therefore, it would be correct to call this kind of nonlocality: 'weak nonlocality'. The related violation of local realism inequalities in SRT, as well as in quantum mechanics, clearly follows from the formalism, indicating a certain parallel between the two theories.

**Acknowledgments.** We are grateful to Dmitriy A. Balakin, Nataliya M. Kolyupanova and Vladimir B. Lapshin for their assistance in the work and useful discussions

**References**

1. A. Einstein, B. Podolsky, and N. Rosen. Phys. Rev. **47**, 777. **DOI**: 10.1103/PhysRev.47.777
2. V.A. Fock, A. Einstein, B. Podolsky, and N. Rosen, N. Bohr. Physics Uspekhi, **XVI**, No 4, 436-457. (1936).
3. A.Yu. Khrennikov. EPR-Bohm experiment and Bell's inequality: Quantum Physics and probability theory. TMF, **157**, No 1, 99–115. (2008). **DOI**: 10.4213/tmf6266

4. N.N. Vorobyov. Theory of Probability and its Applications. **VII**, 2, 153-169 (1962). (In Russion).

5. J.S. Bell. Physics Physique **1**(3) 195 (1964). **DOI**: 10.1103/PhysicsPhysiqueFizika.1.195

6. J.F. Clauser, M.A. Horne, A. Shimony, R.A. Holt. Phys. Rev. Lett. **23**, 880 (1969). **DOI**: 10.1103/PhysRevLett.23.880

7. A.V. Belinskii, D.N. Klyshko. Physics Uspekhi **36** No.8 653 (1993). **DOI**: 10.3367/UFNr.0163.199308a.0001

8. A.V. Belinsky, D.N. Klyshko. Phys.Lett.A **176** No.6 415 (1993).

9. A.V. Belinskii. Physics Uspekhi **37** No.4 413 (1994).

10. L.D. Landau, E.M. Lifshitz, The Classical Theory of Fields. Course of Theoretical Physics. Vol. 2 (4th ed.).

11. N.G. Chetayev. Theoretic Mechanics. Moscow. Nauka. (1987). (In Russian).